\documentclass[prl,twocolumn]{revtex4-1}%
\usepackage{amsmath,amsfonts,amssymb}
\usepackage{xcolor}
\usepackage{graphicx}
\usepackage[T1]{fontenc}
\usepackage[english]{babel}
\usepackage{amsmath}
\usepackage{amsfonts}
\usepackage{amssymb}%
\providecommand{\U}[1]{\protect\rule{.1in}{.1in}}
\pdfoutput=1
\graphicspath{{C:/Users/Jiri/Dropbox/Papers/Physics_papers/2014/Fractals/v15/figures/}{C:/Users/Jiri/Dropbox/Papers/Physics_papers/2014/Fractals/Versions/v22/figures_main/}{./figures_main/}}
\begin{document}
\title{Spectral properties of electrons in fractal nanowires}
\author{Alberto Hernando}
\email{alberto.hernandodecastro@epfl.ch}
\author{Miroslav \v{S}ulc}
\author{Ji\v{r}\' i Van\'i\v{c}ek}
\email{jiri.vanicek@epfl.ch}
\affiliation{Laboratory of Theoretical Physical Chemistry, Institut des Sciences et
Ing\'enierie Chimiques, \'Ecole Polytechnique F\'ed\'erale de Lausanne,
CH-1015 Lausanne, Switzerland}
\date{\today }

\begin{abstract}
In view of promising applications of fractal nanostructures, we analyze the
spectra of quantum particles in the Sierpinski carpet and study the
non-correlated electron gas in this geometry. We show that the spectrum
exhibits scale invariance with almost arbitrary spacing between energy levels,
including large energy gaps at high energies. These features disappear in the
analogous random fractal---where Anderson localization dominates---and in the
regular lattice of equally sized holes---where only two length scales are
present. The fractal structure amplifies microscopic effects, resulting in the
presence of quantum behavior of the electron gas even at high temperatures.
Our results demonstrate the potential of fractal nanostructures to improve the
light-matter interaction at any frequency, with possible applications, e.g.,
in the development of solar cells with a wide absorption spectrum, artificial
photosynthesis, or nanometamaterials with tailored Fermi levels and band gaps,
operating in a wide range of frequencies, and with extended operating
temperature range.

\end{abstract}
\maketitle

New generation of materials, based on nanostructures, is showing its potential
to revolutionise industry \cite{book_Cao,Stangl:2004}. In parallel,
self-similarity properties of fractal structures have revised the way we
manufacture antennas \cite{Hohlfeld:1999} and permitted the production of
novel metamaterials \cite{Huang:2010,Cohen:2012}. New materials and
manufacturing techniques have enabled producing fractal circuits
\cite{Fairbanks:2010}, porous media such as silicas or aerogels, where
superfluid helium flows in a fractal environment \cite{Pollanen:2012},
fractal-shaped clusters of nanoparticles formed by aggregation
\cite{Batabyal:2013}; or even fractal snowflakes composed of graphene
\cite{Massicotte:2013}. Additionally, one can find self-similarity in the
electronic spectra of quantum systems, such as in Hofstadter's butterfly and
in the fractal quantum Hall effect \cite{Hofstadter:1976,Hunt:2013}. Indeed,
quantum fractals are drawing an increasing attention due to their intriguing
properties and potential applications \cite{Katomeris_Evangelou:1996}.

To describe the properties of a quantum particle in a fractal geometry (e.g.,
an electron trapped in a fractal-shaped nanowire), we first find the
eigenstates of the particle's Hamiltonian $\mathcal{H=T+V}$, where
$\mathcal{T}$ is the kinetic energy operator and $\mathcal{V}$ is the
potential energy of an infinite well with the desired fractal boundaries. For
the well, we have chosen the paradigmatic Sierpinski carpet, a fractal object
studied in many different contexts \cite{Volpe:2011,Fairbanks:2010}. This
carpet is constructed by subdividing a square into smaller copies of itself
and by removing the central copy iteratively, as shown in Fig.~1. Each
iteration is labeled with an integer $G$ indicating the \textit{generation} of
the carpet. We study generations up to $G=6$, starting from $G=0$, which
represents the square well. Quantum Hamiltonian for a particle of mass $m$ in
a fractal geometry can be written as $\mathcal{H}=\mathcal{T+V}=-\frac
{\hbar^{2}}{2m}\nabla^{2}+V({\mathbf{q}})$, where $V({\mathbf{q}})$ is an
infinite well potential with the shape of the accessible region in the carpet
of generation $G$ and side $L$. Without loss of generality, units of $L$ for
length, $\hbar^{2}/mL^{2}$ for energy and $k_{B}\hbar^{2}/mL^{2}$ for
temperature (where $k_{B}$ is the Boltzmann constant) are used when $m$,
$\hbar$, or $L$ are not shown explicitly. For any finite generation $G$,
Sierpinski carpet can be viewed as a polygonal billiard with rational angles,
and therefore is an example of pseudointegrable systems discovered by Richens
and Berry \cite{Richens_Berry:1981}, who also studied the equivalent of the
first generation of Sierpinski carpet, albeit with periodic boundary
conditions. As for higher fractal generations, Katomeris and Evangelou studied
properties of Sierpinski \textit{lattices} \cite{Katomeris_Evangelou:1996}; we
instead focus on the dynamics of quantum particles in a fractal Sierpinski
\textit{volume}, such as electrons in fractal shaped wires, which are not yet
well understood.

To obtain the eigenenergies and eigenstates of $\mathcal{H}$, we have used two
different diagonalization methods: the finite differences method (FDM) and
imaginary-time nonuniform-mesh method (ITNUMM)~\cite{Hernando:2013}---see
section I in Supplementary Material (SM) for details. The relative error in
the eigenenergies computed with FDM grows linearly with the energy, whereas
the error remains roughly constant for the ITNUMM. Thus, despite the error of
the FDM being smaller than the error of ITNUMM for the lowest-energy states,
it eventually becomes larger for high-energy states. As a consequence, ITNUMM
is the preferred method if one is interested in energies of many states.
Nevertheless, we have verified all our numerical results by evaluating the
energies both with ITNUMM and FDM.

\begin{figure}
[t]\includegraphics[width=\linewidth]{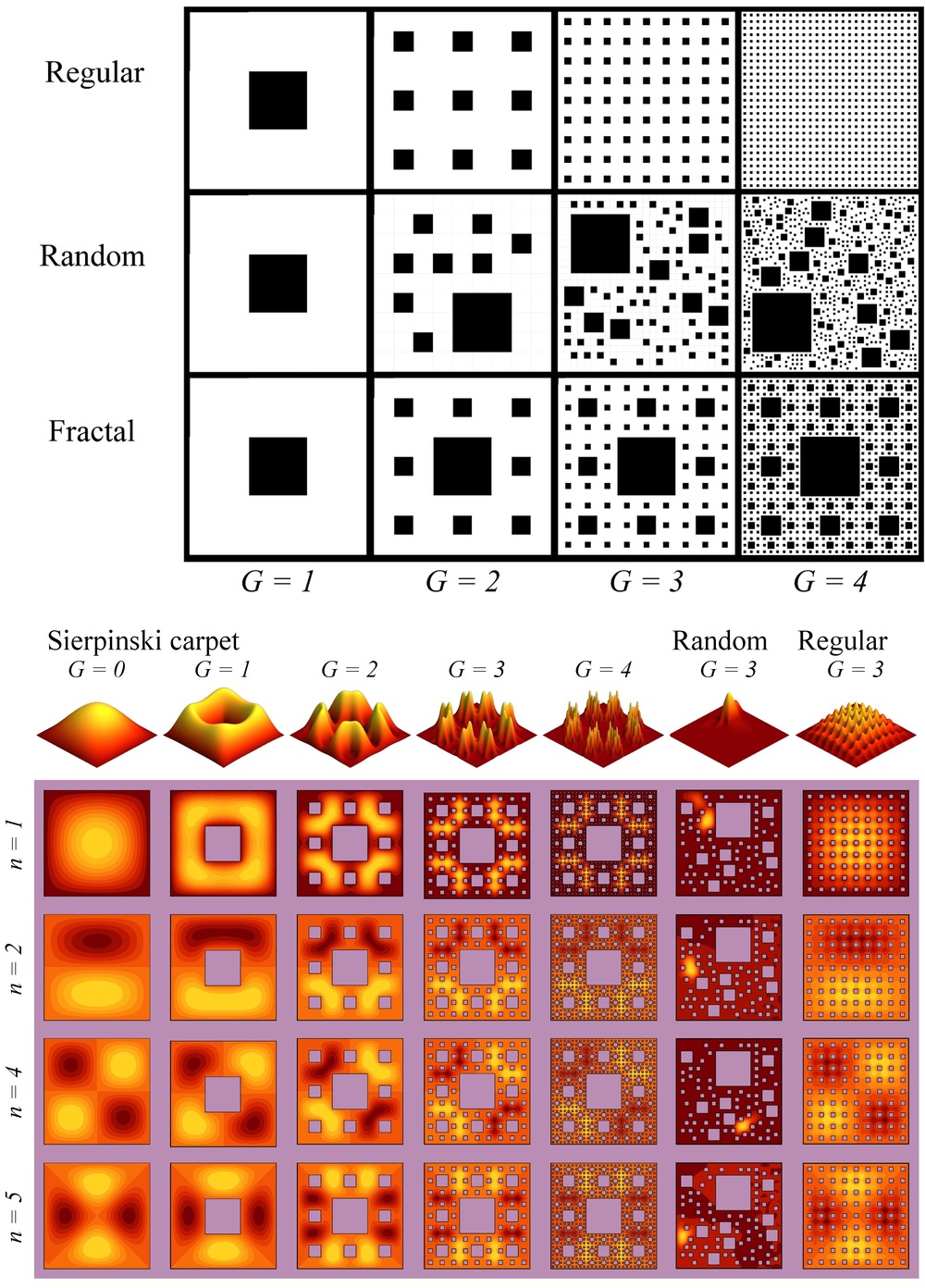}

\caption{\label{fig1}Top panel: Iterative procedure for constructing each generation $G$
of the Sierpinski carpet (``Fractal''), and of the analogous random fractal (``Random'') and regular
non-fractal (``Regular'') structures. Bottom panel:
Wavefunctions of the first five states (rows $n=1$, 2, 4, 5) for
the particle in a box ($G=0$), and for the first four generations of the
Sierpinski carpet ($G=1$, 2, 3, 4). For comparison, corresponding results are
shown for the third generation ($G=3$) of both the random fractal carpet
(where Anderson localization takes place) and the regular non-fractal
lattice.}
\end{figure}

\textit{The spinless single-particle spectrum.} Figure 1 shows the structure
of the lowest-lying states from the zeroth to fourth generation ($G=0$, $1$,
$2$, $3$ and $4$. Generations $G=5$ and $6$ are not shown due to the
limitations of image resolution.) As the fractal generation increases, the
wavefunctions develop intricate structures on smaller and smaller scales,
resulting in the exponential increase of the kinetic energy of the ground
state $\epsilon_{\text{gs}}$ (Fig.~2a). This contribution to the energy
affects equally also all excited states and manifests itself as a global shift
of the energy spectrum for a given generation. Hence, the mode number function
$\mathcal{N}$ (Fig.~2b) and energy levels $\epsilon$ (Fig.~2c) are comparable
for all generations when referred to the ground state. For $G=0$, one obtains
the spectrum of a particle in a two-dimensional well \cite{book_Girifalco},
where the mode number is proportional to the energy as $\mathcal{N}%
(\epsilon)=\epsilon~mL^{2}/2\pi\hbar^{2}$, as predicted by Weyl's law.
Increasing the fractal generation gives rise to two qualitatively different
energy regimes: (i) At lower energies, some levels attract each other,
agglomerating into bands, while others exhibit repulsion, generating large
gaps between the bands; (ii) at higher energies, the mode number eventually
approaches the prediction of Weyl's law $\mathcal{N}(\epsilon)=\epsilon
~m^{\ast}L^{2}/2\pi\hbar^{2}$, where $m^{\ast}=(8/9)^{G}m$ is the effective
mass. Using this analytical form to rescale mode numbers as $\mathcal{N}%
^{\ast}=\mathcal{N}/8^{G}$ and energies as $\epsilon^{\ast}=\epsilon
mL^{2}/9^{G}2\pi\hbar^{2}$ (see Fig. 2d, left panel), we find, remarkably,
that \textit{for any generation }$G$ the bands and gaps in regime (i) appear
in the part of the spectrum with energies $\epsilon^{\ast}<1$ ($\mathcal{N}%
^{\ast}<1$), whereas the approximately constant density of states in regime
(ii) appears for energies $\epsilon^{\ast}>1$ ($\mathcal{N}^{\ast}>1$). This
scaling of the energy levels demonstrates that the self-similarity in the
geometry is reflected also in a scale-invariant spectrum.

\begin{figure}
[t]\includegraphics[width=0.9\linewidth]{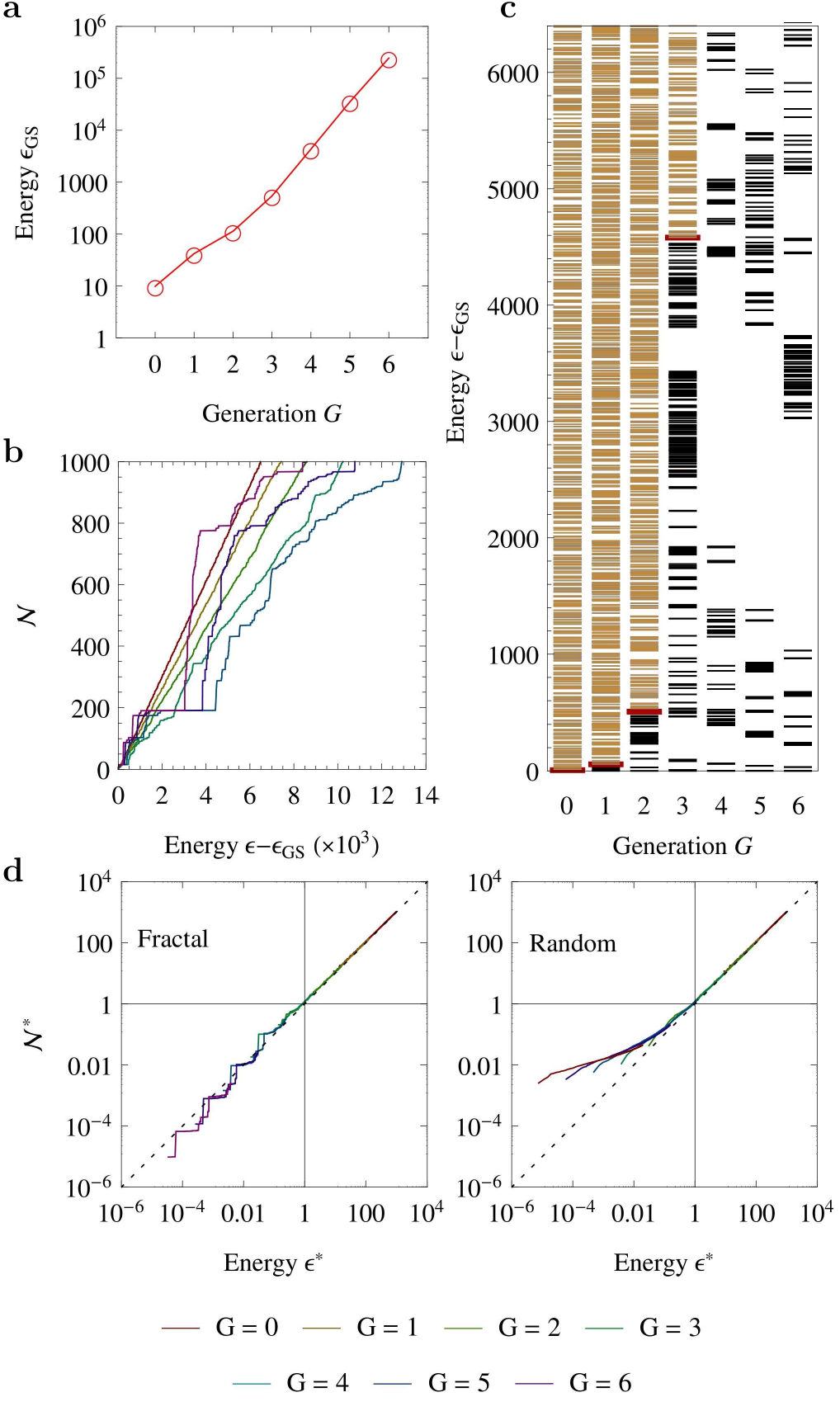}
\caption{Spectrum of a single particle in a Sierpinski-shaped well. \textbf{a,} Ground state energy $\protect\epsilon_{\text{gs}}$: The exponential
decrease of the smallest length scale with increasing generation $G$
manifests itself in the exponential increase of $\protect\epsilon_{\text{gs}} $. \textbf{b,} Mode number function $\mathcal{N}(\protect\epsilon-\protect\epsilon_{\text{gs}})$. \textbf{c,} Energy level diagram: The border between
low-energy, band-gap regime (black line segments) and the almost continuous
high-energy regime (orange line segments) satisfying Weyl's law (see text)
is shown for each $G$ by a thick red line segment. \textbf{d,} The
self-similarity of the spectrum becomes apparent when the energies and
levels are rescaled according to Weyl's law (left panel: Sierpinski carpet;
right panel: random carpet).} \label{fig3}
\end{figure}

To further explore this phenomenon, we studied the eigenvalue nearest-neighbor
spacing distribution in both regimes for all six generations. \ \ This
distribution is widely used in the fields of random matrix theory and quantum
chaos for characterizing ensembles of random matrices and dynamical properties
of Hamiltonian systems \cite{book_Reichl,Edelman:2005}. In addition, since the
energy of any optical transition of an electron can be described as a sum of
level spacings, their distribution can provide some information about the
expected absorption/emission properties of fractal nanowires. After first
unfolding the energy levels (as described in section II of the SM), we
computed the spacings $s$ of every two consecutive modes as well as the
maximum likelihood estimates of the parameters $\alpha$ and $b$ of the Brody
distribution \cite{Brody:1981}
\begin{equation}
p(s)=(1+b)\alpha s^{b}\exp\left(  -\alpha s^{b+1}\right)  \label{eqBD}%
\end{equation}
fit to the probability density of $s$. Whereas Brody distribution with zero
Brody parameter ($b=0$) reduces to the exponential distribution $p(s)=\alpha
\exp\left(  -\alpha s\right)  $, reflecting Poisson level statistics
characteristic of \emph{generic integrable} systems \cite{Berry_Tabor:1977},
unit Brody parameter ($b=1$) yields the Wigner distribution $p(s)=2\alpha
s\exp\left(  -\alpha s^{2}\right)  $ and, according to the quantum-chaos
conjecture
\cite{McDonald_Kaufman:1979,Berry:1981,Bohigas_Schmit:1984,Seligman_Zirnbauer:1984}%
, reflects classically \emph{chaotic} dynamics of systems with time-reversal
invariant Hamiltonian. Generic time-reversal invariant systems with mixed
dynamics and without symmetry can be well described with a mixture of
exponential and Wigner distributions
\cite{Seligman_Zirnbauer:1984,Berry_Robnik:1984}; while this mixture itself is
not a Brody distribution, it can be typically well fitted with a Brody
distribution with $0<b<1$. Remarkably, as shown in Fig.~3, for the Sierpinski
carpet we find that the Brody parameter evolves from $b=-0.03\pm0.07$ for
$G=1$, corresponding to Poisson level statistics ($b=0$), to $b=-0.96\pm0.03$
for $G\geq6$ in regime (i), resulting in almost an inverse power law
distribution ($b=-1$). As the generation increases, the level
repulsion-attraction intensifies and it becomes harder to define the mean
level spacing. Indeed, for the asymptotic value $b=-1$ the mean level spacing
ceases to exist and the distribution cannot even be normalized; spacings of
any order of magnitude are equally probable at any energy. {This transition
does not appear in regime (ii), where Brody parameter $b\approx0$ in all the
cases studied here. While transitions between non-negative Brody parameters
(from $b=1$ to $b=0$) have been seen in the family of Koch fractals of
changing fractal dimension \cite{Sakhr:2005}, to the best of our knowledge a
transition from Poisson statistics to a long-tailed distribution, similar to
the one described here, was previously observed only by Katomeris and
Evangelou, who saw a power law distribution with exponent }$-1.56${\ in a
fractal lattice by removing diagonal disorder \cite{Katomeris_Evangelou:1996}.
However, negative Brody parameters were observed previously in superintegrable
systems ($b\approx-0.24$) \cite{Alhassid_Whelan:1990} or 1D lattices with
unbounded quantum diffusion ($b=-3/2)$ \cite{Geisel_Petschel:1991}. For the
desymmetrized system (see SM section II), we find a qualitatively similar
behavior involving a transition from positive to negative values, confirming
that the effect is due to the fractal geometry and not due to degeneracies
from simple geometrical symmetries. }

\begin{figure}
[t]\includegraphics[width=\linewidth]{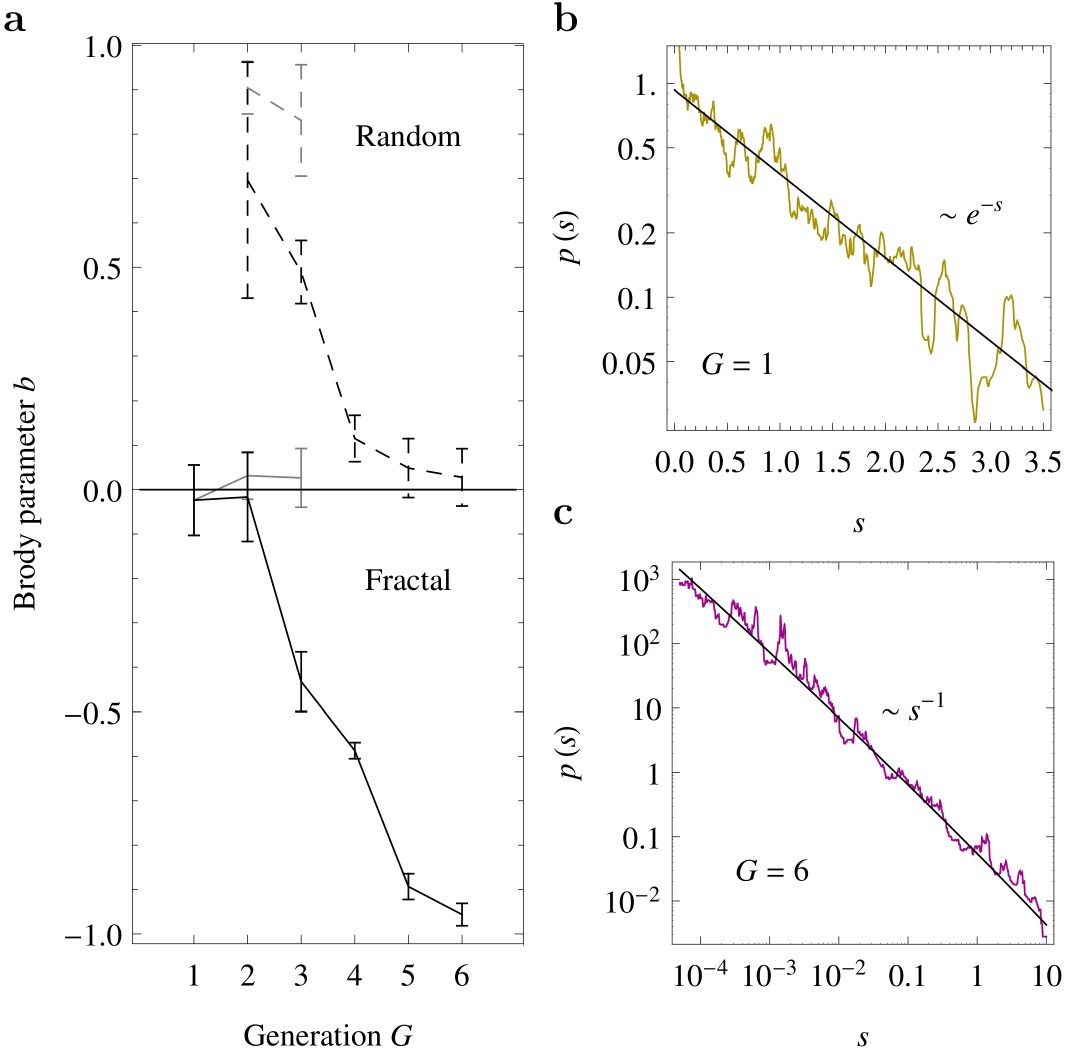}

\caption{ \textbf{a,} dependence of the Brody parameter $b$ of the
distribution of level spacings on the fractal generation $G$. Solid black: the
low-energy regime (i) (see text) undergoes a transition from the exponential
(\textbf{b,} for $G=1$, $b\approx0$ reflecting Poisson level statistics) to inverse power-law distribution
(\textbf{c,} for $G \geq 6$, $b\approx-1$); Solid gray: the high-energy regime (ii) shows
no (or at most weak) dependence on $G$, approximately following the Poisson distribution for all $G$. Dashed black: low energy regime (i) for the random fractal, evolving from
the Wigner ($b\approx1$) to exponential ($b\approx0$) distribution; Dashed gray: regime (ii) showing no (or at most weak) dependence on $G \geq 2$.}
\end{figure}

This peculiar feature of the spectrum of the Sierpinski carpet disappears
(Fig.~3 and SM Fig.~9) both in the analogous random fractal carpet, where the
holes are distributed randomly without overlapping (Fig.~1), and in the
equivalent regular but non-fractal geometry, where all the holes are of the
same size and are distributed regularly in a lattice (Fig.~1).

The random fractal exhibits a quasi-continuous spectrum instead of the level
repulsion and attraction pattern. Even if the potential well exhibits scale
invariance in a statistical fashion, breaking the geometrical symmetries
results in Anderson localization \cite{Anderson:1958}, typical of disordered
media and here reflected in the localized eigenfunctions in Fig.~1. After
rescaling the mode numbers and energies as for the fractal, we find the same
regime at high energies $\epsilon^{\ast}>1$ ($\mathcal{N}^{\ast}>1$, Fig.~2d
right panel). However, in this regime for $G=2$ and $3$ the Brody parameter is
$b=0.90\pm0.06$ and $b=0.83\pm0.12$ respectively (Fig.~3), close to the value
$1$ characteristic of chaotic systems. On the other hand, the low-energy part
of the spectrum---where Anderson states are found---lacks the band and gap
structure, and the Brody parameter evolves, remarkably, from $b=0.7\pm0.2$ for
$G=2$ to $b=0.03\pm0.06$ for $G=6$ as for an integrable system (see section II
of SM for details). Thus, breaking the regularity of the fractal results in
the loss of the scale invariance of nearest-neighbor level spacings.

As another control, we analyzed the equivalent regular lattice of holes of
equal size. This system retains the geometrical symmetries of the square but
lacks the scale invariance since it has only two characteristic scales---the
size of the holes and the size $L$ of the system. We find that the spectrum of
the lattice exhibits none of the remarkable properties of the fractal (scaling
of the energy levels, presence of a regime with band-gap structure,
self-similarity). This system must be desymmetrized in order to fit the
spacings distribution to the form given by Eq.~(\ref{eqBD}); the fitted
distribution has a positive Brody parameter that evolves to $b\approx0$ for
the largest generations (see SM section II). \textit{Overall, these two
negative controls suggest that the unique features of the spectrum of the
fractal are due to the combined effect of geometrical symmetries and scale
invariance}.

\textit{The electron gas.} We used 1000 single-particle states in order to
study the thermodynamic properties of the non-correlated electron gas
\cite{book_Girifalco} in a Sierpinski carpet. The Fermi level of the system
for different total number of electrons $N_{e}$ with spin degeneracy is
presented in Fig.~4. It is shown that the generation $G$ can be chosen to
obtain a Fermi level located close to a gap, and reproduce the properties of a
semiconductor material with tailored band gap. The heat capacity $C(T)$ for
$N_{e}=100$ electrons is shown in Fig.~4 (see section III of SM for details).
The 1000 states suffice to describe each system accurately in a window of
temperatures of almost three times the Fermi temperature $T_{F}$ ($T_{F}=\pi
N_{F}\approx314$ for $G=0$). For $G=0$ we recover the behavior of the 2D
electron gas (2DEG), as expected. Indeed, $C(T)$ depends linearly on $T$ at
low temperatures and converges to the classical limit ($C_{cl}=N_{e}$ in two
dimensions) at $T\sim2T_{F}$. For $G=1$ and $G=2$ we obtain a similar
behavior, with small deviations from the behavior of the 2DEG. However, for
$G\geq3$ the presence of quasi-continuous bands separated by large energy gaps
becomes important at low temperatures, where the increasing temperature
permits some electrons to suddenly cross the energy gaps and populate higher
energy bands in a process that is reflected in the oscillations of the heat
capacity as a function of temperature. For the highest generations,
oscillations become substantial in a wider range of temperatures and, at
higher temperatures, the heat capacity eventually overshoots the classical
limit. Remarkably, the dependence of the heat capacity per particle on the
temperature expressed in units of Fermi temperature converges to the same
limit curve for the highest generations (Fig. 4).

\begin{figure}
\centering\includegraphics[width=\columnwidth]{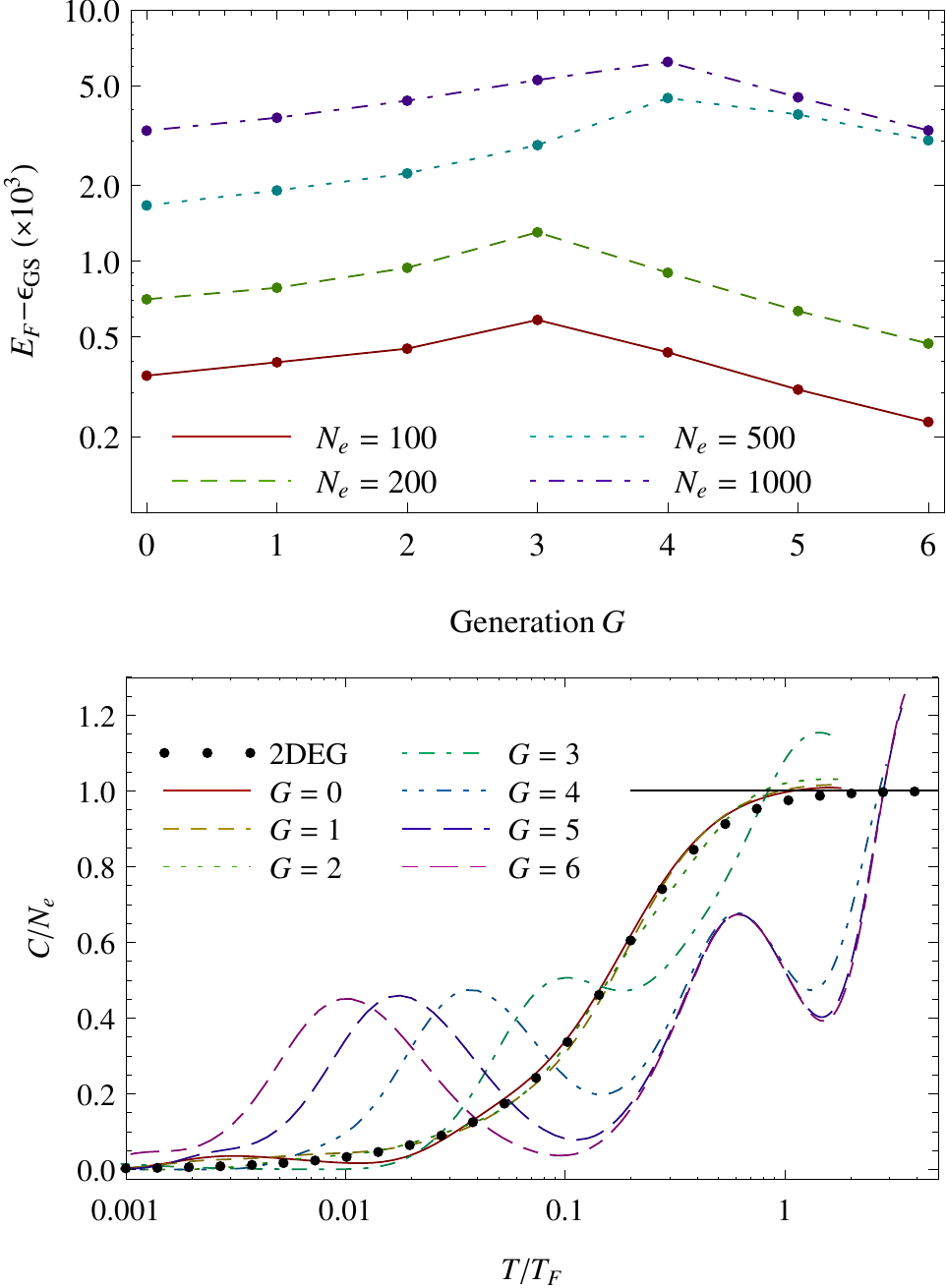} \caption{
Top panel: Fermi energy $E_F$ with respect the ground state energy
of noninteracting electron gas for $N_{e}=100$, $200$, $500$, and $1000$
electrons for all generations studied here.
Bottom panel: Dependence of the heat capacity per particle $C/N_e$ on temperature
(in units of Fermi Temperature) $T/T_F$ for
$N_{e}=100$ electrons. The alternation of bands and gaps
generates oscillations in $C$ as electrons thermally populate higher energy bands, and results in an overshooting with respect to the
asymptotic thermodynamical value $C=N_{e}$ for a 2DEG (shown as a horizontal black line),
even at temperatures where the equivalent non-fractal 2DEG has reached the
classical limit. Remarkably, it converges to a well-defined function of $T/T_F$ as the generation increases.}
\end{figure}

Two different mechanisms contribute to the overshooting: On one hand, the
observed oscillations due to the thermal population of high energy bands in
regime (i) become more relevant as the generation increases. Figure~4 shows
that for $N_{e}=100$ at a temperature of $T\sim2T_{F}$ where the 2DEG ($G=0$)
already behaves classically, an analogous system of the same size but with the
shape of a Sierpinski carpet still behaves quantum-mechanically. On the other
hand, the increasing value of the ratio of the perimeter and surface of the
fractal with increasing $G$ affects the energy dependence of the density of
states in regime (ii). Indeed, the first correction term to Weyl's law is
proportional to the perimeter and its dependence on $\sqrt{\epsilon}$ gives
rise to the overshooting (see SM section IV), this effect becoming more
pronounced as the perimeter term dominates over the surface term. However, for
any finite value of $G$, the surface contribution eventually dominates over
the perimeter correction at sufficiently high energies, and thus the
asymptotic value agrees with the 2D ideal-gas value. This would \textit{not}
happen in the limit $G\rightarrow\infty$ (for a fixed energy or temperature);
in this limit the surface of the fractal becomes zero while the perimeter
becomes infinite, yielding a non-classical result of a divergent heat capacity.

In summary, the energy spectrum of quantum particles trapped in a Sierpinski
structure shows self-similarity, with level attraction and repulsion leading
to bands and gaps at any scale. The distribution of consecutive energy level
spacings exhibits a transition from an exponential to inverse power law
density as the fractal generation increases. At high generations, no
characteristic size for the spacings can be defined and level spacings of any
order of magnitude are equally probable. In fractal geometries, quantum
effects originating from the discreteness of the energy levels can be observed
at temperatures where the equivalent non-fractal system already behaves classically.

In view of our results, fractal-shaped wires could exhibit promising features
for a wide range of applications. Scale invariance of the spectrum could
improve the photoelectric interaction of nanodevices in a wide range of light
frequencies and allow the design of nanomaterials with tailored Fermi levels,
band gaps, and conduction bands that can be thermally populated---mimicking
the properties of semiconductors, as seen in the oscillations of the heat
capacity---with potential applications for solar cells
\cite{Zhu:2013,Abdellatif:2013}, artificial photosynthesis \cite{Wang:2009},
or metamaterials \cite{Huang:2010,Cohen:2012,Volpe:2011,Schurig:2006}. In
addition, the persistence of quantum effects at high temperatures could
increase the operating temperature range of these quantum nanodevices.

\begin{acknowledgments}
The authors would like to thank Sir Michael Berry, Rub\'{e}n Fossil\'{o}n, Lev
Kaplan, Angelo Plastino, Thomas Seligman, Marius Wehrle, and Eduardo Zambrano
for useful discussions and valuable comments. This research was supported by
the Swiss National Science Foundation (NSF(CH)) with Grant No.
200020\textunderscore{}150098 and National Center of Competence in Research
(NCCR) Molecular Ultrafast Science and Technology (MUST), and by the EPFL.
\end{acknowledgments}

\bibliographystyle{apsrev4-1}

\begin{thebibliography}{33}%
\makeatletter
\providecommand \@ifxundefined [1]{%
 \@ifx{#1\undefined}
}%
\providecommand \@ifnum [1]{%
 \ifnum #1\expandafter \@firstoftwo
 \else \expandafter \@secondoftwo
 \fi
}%
\providecommand \@ifx [1]{%
 \ifx #1\expandafter \@firstoftwo
 \else \expandafter \@secondoftwo
 \fi
}%
\providecommand \natexlab [1]{#1}%
\providecommand \enquote  [1]{``#1''}%
\providecommand \bibnamefont  [1]{#1}%
\providecommand \bibfnamefont [1]{#1}%
\providecommand \citenamefont [1]{#1}%
\providecommand \href@noop [0]{\@secondoftwo}%
\providecommand \href [0]{\begingroup \@sanitize@url \@href}%
\providecommand \@href[1]{\@@startlink{#1}\@@href}%
\providecommand \@@href[1]{\endgroup#1\@@endlink}%
\providecommand \@sanitize@url [0]{\catcode `\\12\catcode `\$12\catcode
  `\&12\catcode `\#12\catcode `\^12\catcode `\_12\catcode `\%12\relax}%
\providecommand \@@startlink[1]{}%
\providecommand \@@endlink[0]{}%
\providecommand \url  [0]{\begingroup\@sanitize@url \@url }%
\providecommand \@url [1]{\endgroup\@href {#1}{\urlprefix }}%
\providecommand \urlprefix  [0]{URL }%
\providecommand \Eprint [0]{\href }%
\providecommand \doibase [0]{http://dx.doi.org/}%
\providecommand \selectlanguage [0]{\@gobble}%
\providecommand \bibinfo  [0]{\@secondoftwo}%
\providecommand \bibfield  [0]{\@secondoftwo}%
\providecommand \translation [1]{[#1]}%
\providecommand \BibitemOpen [0]{}%
\providecommand \bibitemStop [0]{}%
\providecommand \bibitemNoStop [0]{.\EOS\space}%
\providecommand \EOS [0]{\spacefactor3000\relax}%
\providecommand \BibitemShut  [1]{\csname bibitem#1\endcsname}%
\let\auto@bib@innerbib\@empty
\bibitem [{\citenamefont {Cao}\ and\ \citenamefont {Wang}(2011)}]{book_Cao}%
  \BibitemOpen
  \bibfield  {author} {\bibinfo {author} {\bibfnamefont {G.}~\bibnamefont
  {Cao}}\ and\ \bibinfo {author} {\bibfnamefont {Y.}~\bibnamefont {Wang}},\
  }in\ \href@noop {} {\emph {\bibinfo {booktitle} {World Scientific Series in
  Nanoscience and Nanotechnology: Volume 2}}},\ \bibinfo {editor} {edited by\
  \bibinfo {editor} {\bibfnamefont {F.}~\bibnamefont {Spaepen}}}\ (\bibinfo
  {publisher} {{World Scientific}},\ \bibinfo {address} {Singapore},\ \bibinfo
  {year} {2011})\ \bibinfo {edition} {2nd}\ ed.\BibitemShut {Stop}%
\bibitem [{\citenamefont {Stangl}\ \emph {et~al.}(2004)\citenamefont {Stangl},
  \citenamefont {Holy},\ and\ \citenamefont {Bauer}}]{Stangl:2004}%
  \BibitemOpen
  \bibfield  {author} {\bibinfo {author} {\bibfnamefont {J.}~\bibnamefont
  {Stangl}}, \bibinfo {author} {\bibfnamefont {V.}~\bibnamefont {Holy}}, \ and\
  \bibinfo {author} {\bibfnamefont {G.}~\bibnamefont {Bauer}},\ }\href@noop {}
  {\bibfield  {journal} {\bibinfo  {journal} {Rev. Mod. Phys.}\ }\textbf
  {\bibinfo {volume} {76}},\ \bibinfo {pages} {725} (\bibinfo {year}
  {2004})}\BibitemShut {NoStop}%
\bibitem [{\citenamefont {Hohlfeld}\ and\ \citenamefont
  {Cohen}(1999)}]{Hohlfeld:1999}%
  \BibitemOpen
  \bibfield  {author} {\bibinfo {author} {\bibfnamefont {R.}~\bibnamefont
  {Hohlfeld}}\ and\ \bibinfo {author} {\bibfnamefont {N.}~\bibnamefont
  {Cohen}},\ }\href@noop {} {\bibfield  {journal} {\bibinfo  {journal}
  {Fractals}\ }\textbf {\bibinfo {volume} {7}},\ \bibinfo {pages} {79}
  (\bibinfo {year} {1999})}\BibitemShut {NoStop}%
\bibitem [{\citenamefont {Huang}(2010)}]{Huang:2010}%
  \BibitemOpen
  \bibfield  {author} {\bibinfo {author} {\bibfnamefont {X.}~\bibnamefont
  {Huang}},\ }\href@noop {} {\bibfield  {journal} {\bibinfo  {journal} {Opt.
  Express}\ }\textbf {\bibinfo {volume} {18}},\ \bibinfo {pages} {10377}
  (\bibinfo {year} {2010})}\BibitemShut {NoStop}%
\bibitem [{\citenamefont {Cohen}(2012)}]{Cohen:2012}%
  \BibitemOpen
  \bibfield  {author} {\bibinfo {author} {\bibfnamefont {N.}~\bibnamefont
  {Cohen}},\ }\href@noop {} {\bibfield  {journal} {\bibinfo  {journal}
  {Fractals}\ }\textbf {\bibinfo {volume} {20}},\ \bibinfo {pages} {227}
  (\bibinfo {year} {2012})}\BibitemShut {NoStop}%
\bibitem [{\citenamefont {Fairbanks}(2010)}]{Fairbanks:2010}%
  \BibitemOpen
  \bibfield  {author} {\bibinfo {author} {\bibfnamefont {M.}~\bibnamefont
  {Fairbanks}},\ }\emph {\bibinfo {title} {Transposrt in micro to nanoscale
  solid state networks}},\ \href@noop {} {Ph.D. thesis},\ \bibinfo  {school}
  {University of Oregon} (\bibinfo {year} {2010})\BibitemShut {NoStop}%
\bibitem [{\citenamefont {Pollanen}\ \emph {et~al.}(2012)\citenamefont
  {Pollanen}, \citenamefont {Li}, \citenamefont {Collet}, \citenamefont
  {Gannon}, \citenamefont {Halperin},\ and\ \citenamefont
  {Sauls}}]{Pollanen:2012}%
  \BibitemOpen
  \bibfield  {author} {\bibinfo {author} {\bibfnamefont {J.}~\bibnamefont
  {Pollanen}}, \bibinfo {author} {\bibfnamefont {J.}~\bibnamefont {Li}},
  \bibinfo {author} {\bibfnamefont {C.}~\bibnamefont {Collet}}, \bibinfo
  {author} {\bibfnamefont {W.}~\bibnamefont {Gannon}}, \bibinfo {author}
  {\bibfnamefont {W.}~\bibnamefont {Halperin}}, \ and\ \bibinfo {author}
  {\bibfnamefont {J.}~\bibnamefont {Sauls}},\ }\href@noop {} {\bibfield
  {journal} {\bibinfo  {journal} {Nature Phys.}\ }\textbf {\bibinfo {volume}
  {8}},\ \bibinfo {pages} {317} (\bibinfo {year} {2012})}\BibitemShut {NoStop}%
\bibitem [{\citenamefont {Batabyal.}(2013)}]{Batabyal:2013}%
  \BibitemOpen
  \bibfield  {author} {\bibinfo {author} {\bibfnamefont {R.}~\bibnamefont
  {Batabyal.}},\ }\href@noop {} {\bibfield  {journal} {\bibinfo  {journal} {J.
  Appl. Phys.}\ }\textbf {\bibinfo {volume} {114}},\ \bibinfo {pages} {064304}
  (\bibinfo {year} {2013})}\BibitemShut {NoStop}%
\bibitem [{\citenamefont {Massicotte.}(2013)}]{Massicotte:2013}%
  \BibitemOpen
  \bibfield  {author} {\bibinfo {author} {\bibfnamefont {M.}~\bibnamefont
  {Massicotte.}},\ }\href@noop {} {\bibfield  {journal} {\bibinfo  {journal}
  {Nanotechnology}\ }\textbf {\bibinfo {volume} {24}},\ \bibinfo {pages}
  {325601} (\bibinfo {year} {2013})}\BibitemShut {NoStop}%
\bibitem [{\citenamefont {Hofstadter}(1976)}]{Hofstadter:1976}%
  \BibitemOpen
  \bibfield  {author} {\bibinfo {author} {\bibfnamefont {D.}~\bibnamefont
  {Hofstadter}},\ }\href@noop {} {\bibfield  {journal} {\bibinfo  {journal}
  {Phys.\ Rev.~B}\ }\textbf {\bibinfo {volume} {14}},\ \bibinfo {pages} {2239}
  (\bibinfo {year} {1976})}\BibitemShut {NoStop}%
\bibitem [{\citenamefont {et~al.}(2013)}]{Hunt:2013}%
  \BibitemOpen
  \bibfield  {author} {\bibinfo {author} {\bibfnamefont {B.~H.}\ \bibnamefont
  {et~al.}},\ }\href@noop {} {\bibfield  {journal} {\bibinfo  {journal}
  {Science}\ }\textbf {\bibinfo {volume} {340}},\ \bibinfo {pages} {1427}
  (\bibinfo {year} {2013})}\BibitemShut {NoStop}%
\bibitem [{\citenamefont {Katomeris}\ and\ \citenamefont
  {Evangelou}(1996)}]{Katomeris_Evangelou:1996}%
  \BibitemOpen
  \bibfield  {author} {\bibinfo {author} {\bibfnamefont {G.~N.}\ \bibnamefont
  {Katomeris}}\ and\ \bibinfo {author} {\bibfnamefont {S.~N.}\ \bibnamefont
  {Evangelou}},\ }\href@noop {} {\bibfield  {journal} {\bibinfo  {journal} {J.
  Phys. A: Math. Gen.}\ }\textbf {\bibinfo {volume} {29}},\ \bibinfo {pages}
  {2379} (\bibinfo {year} {1996})}\BibitemShut {NoStop}%
\bibitem [{\citenamefont {Volpe}\ \emph {et~al.}(2011)\citenamefont {Volpe},
  \citenamefont {Volpe},\ and\ \citenamefont {Quidant}}]{Volpe:2011}%
  \BibitemOpen
  \bibfield  {author} {\bibinfo {author} {\bibfnamefont {G.}~\bibnamefont
  {Volpe}}, \bibinfo {author} {\bibfnamefont {G.}~\bibnamefont {Volpe}}, \ and\
  \bibinfo {author} {\bibfnamefont {R.}~\bibnamefont {Quidant}},\ }\href@noop
  {} {\bibfield  {journal} {\bibinfo  {journal} {Opt. Lett.}\ }\textbf
  {\bibinfo {volume} {19}},\ \bibinfo {pages} {3612} (\bibinfo {year}
  {2011})}\BibitemShut {NoStop}%
\bibitem [{\citenamefont {Richens}\ and\ \citenamefont
  {Berry}(1981)}]{Richens_Berry:1981}%
  \BibitemOpen
  \bibfield  {author} {\bibinfo {author} {\bibfnamefont {P.~J.}\ \bibnamefont
  {Richens}}\ and\ \bibinfo {author} {\bibfnamefont {M.~V.}\ \bibnamefont
  {Berry}},\ }\href {\doibase 10.1016/0167-2789(81)90024-5} {\bibfield
  {journal} {\bibinfo  {journal} {Physica D}\ }\textbf {\bibinfo {volume}
  {2}},\ \bibinfo {pages} {495} (\bibinfo {year} {1981})}\BibitemShut {NoStop}%
\bibitem [{\citenamefont {Hernando}\ and\ \citenamefont
  {Van\'{i}\v{c}ek}(2013)}]{Hernando:2013}%
  \BibitemOpen
  \bibfield  {author} {\bibinfo {author} {\bibfnamefont {A.}~\bibnamefont
  {Hernando}}\ and\ \bibinfo {author} {\bibfnamefont {J.}~\bibnamefont
  {Van\'{i}\v{c}ek}},\ }\href@noop {} {\bibfield  {journal} {\bibinfo
  {journal} {Phys.\ Rev.~A}\ }\textbf {\bibinfo {volume} {88}},\ \bibinfo
  {pages} {062107} (\bibinfo {year} {2013})}\BibitemShut {NoStop}%
\bibitem [{\citenamefont {Girifalco}(2000)}]{book_Girifalco}%
  \BibitemOpen
  \bibfield  {author} {\bibinfo {author} {\bibfnamefont {L.}~\bibnamefont
  {Girifalco}},\ }\href@noop {} {\emph {\bibinfo {title} {{Statistical
  Mechanics of Solids}}}}\ (\bibinfo  {publisher} {{Oxford Univ. Press}},\
  \bibinfo {address} {Oxford},\ \bibinfo {year} {2000})\BibitemShut {NoStop}%
\bibitem [{\citenamefont {Reichl}(2004)}]{book_Reichl}%
  \BibitemOpen
  \bibfield  {author} {\bibinfo {author} {\bibfnamefont {L.}~\bibnamefont
  {Reichl}},\ }\href@noop {} {\emph {\bibinfo {title} {{The Transition to
  Chaos: Conservative Classical Systems and Quantum Manifestations}}}},\
  \bibinfo {edition} {{2}}\ ed.\ (\bibinfo  {publisher} {{Springer-Verlag}},\
  \bibinfo {address} {New York},\ \bibinfo {year} {2004})\BibitemShut {NoStop}%
\bibitem [{\citenamefont {Edelman}\ and\ \citenamefont
  {Rao}(2005)}]{Edelman:2005}%
  \BibitemOpen
  \bibfield  {author} {\bibinfo {author} {\bibfnamefont {A.}~\bibnamefont
  {Edelman}}\ and\ \bibinfo {author} {\bibfnamefont {N.}~\bibnamefont {Rao}},\
  }\href@noop {} {\bibfield  {journal} {\bibinfo  {journal} {Acta Numer.}\
  }\textbf {\bibinfo {volume} {14}},\ \bibinfo {pages} {233} (\bibinfo {year}
  {2005})}\BibitemShut {NoStop}%
\bibitem [{\citenamefont {Brody}\ \emph {et~al.}(1981)\citenamefont {Brody},
  \citenamefont {Flores}, \citenamefont {French}, \citenamefont {Mello},
  \citenamefont {Pandey},\ and\ \citenamefont {Wong}}]{Brody:1981}%
  \BibitemOpen
  \bibfield  {author} {\bibinfo {author} {\bibfnamefont {T.~A.}\ \bibnamefont
  {Brody}}, \bibinfo {author} {\bibfnamefont {J.}~\bibnamefont {Flores}},
  \bibinfo {author} {\bibfnamefont {J.~B.}\ \bibnamefont {French}}, \bibinfo
  {author} {\bibfnamefont {P.~A.}\ \bibnamefont {Mello}}, \bibinfo {author}
  {\bibfnamefont {A.}~\bibnamefont {Pandey}}, \ and\ \bibinfo {author}
  {\bibfnamefont {S.~S.~M.}\ \bibnamefont {Wong}},\ }\href {\doibase
  10.1103/RevModPhys.53.385} {\bibfield  {journal} {\bibinfo  {journal} {Rev.
  Mod. Phys.}\ }\textbf {\bibinfo {volume} {53}},\ \bibinfo {pages} {385}
  (\bibinfo {year} {1981})}\BibitemShut {NoStop}%
\bibitem [{\citenamefont {Berry}\ and\ \citenamefont
  {Tabor}(1977)}]{Berry_Tabor:1977}%
  \BibitemOpen
  \bibfield  {author} {\bibinfo {author} {\bibfnamefont {M.~V.}\ \bibnamefont
  {Berry}}\ and\ \bibinfo {author} {\bibfnamefont {M.}~\bibnamefont {Tabor}},\
  }\href {\doibase 10.1098/rspa.1977.0140} {\bibfield  {journal} {\bibinfo
  {journal} {Proc. Royal Soc. London A}\ }\textbf {\bibinfo {volume} {356}},\
  \bibinfo {pages} {375} (\bibinfo {year} {1977})}\BibitemShut {NoStop}%
\bibitem [{\citenamefont {McDonald}\ and\ \citenamefont
  {Kaufman}(1979)}]{McDonald_Kaufman:1979}%
  \BibitemOpen
  \bibfield  {author} {\bibinfo {author} {\bibfnamefont {S.~W.}\ \bibnamefont
  {McDonald}}\ and\ \bibinfo {author} {\bibfnamefont {A.~N.}\ \bibnamefont
  {Kaufman}},\ }\href@noop {} {\bibfield  {journal} {\bibinfo  {journal} {Phys.
  Rev. Lett.}\ }\textbf {\bibinfo {volume} {42}},\ \bibinfo {pages} {1189}
  (\bibinfo {year} {1979})}\BibitemShut {NoStop}%
\bibitem [{\citenamefont {Berry}(1981)}]{Berry:1981}%
  \BibitemOpen
  \bibfield  {author} {\bibinfo {author} {\bibfnamefont {M.~V.}\ \bibnamefont
  {Berry}},\ }\href {\doibase http://dx.doi.org/10.1016/0003-4916(81)90189-5}
  {\bibfield  {journal} {\bibinfo  {journal} {Annals of Phys.}\ }\textbf
  {\bibinfo {volume} {131}},\ \bibinfo {pages} {163} (\bibinfo {year}
  {1981})}\BibitemShut {NoStop}%
\bibitem [{\citenamefont {Bohigas}\ \emph {et~al.}(1984)\citenamefont
  {Bohigas}, \citenamefont {Giannoni},\ and\ \citenamefont
  {Schmit}}]{Bohigas_Schmit:1984}%
  \BibitemOpen
  \bibfield  {author} {\bibinfo {author} {\bibfnamefont {O.}~\bibnamefont
  {Bohigas}}, \bibinfo {author} {\bibfnamefont {M.~J.}\ \bibnamefont
  {Giannoni}}, \ and\ \bibinfo {author} {\bibfnamefont {C.}~\bibnamefont
  {Schmit}},\ }\href {\doibase Doi 10.1103/Physrevlett.52.1} {\bibfield
  {journal} {\bibinfo  {journal} {Phys. Rev. Lett.}\ }\textbf {\bibinfo
  {volume} {52}},\ \bibinfo {pages} {1} (\bibinfo {year} {1984})}\BibitemShut
  {NoStop}%
\bibitem [{\citenamefont {Seligman}\ \emph {et~al.}(1984)\citenamefont
  {Seligman}, \citenamefont {Verbaarschot},\ and\ \citenamefont
  {Zirnbauer}}]{Seligman_Zirnbauer:1984}%
  \BibitemOpen
  \bibfield  {author} {\bibinfo {author} {\bibfnamefont {T.~H.}\ \bibnamefont
  {Seligman}}, \bibinfo {author} {\bibfnamefont {J.~J.~M.}\ \bibnamefont
  {Verbaarschot}}, \ and\ \bibinfo {author} {\bibfnamefont {M.~R.}\
  \bibnamefont {Zirnbauer}},\ }\href@noop {} {\bibfield  {journal} {\bibinfo
  {journal} {Phys. Rev. Lett.}\ }\textbf {\bibinfo {volume} {53}},\ \bibinfo
  {pages} {215} (\bibinfo {year} {1984})}\BibitemShut {NoStop}%
\bibitem [{\citenamefont {Berry}\ and\ \citenamefont
  {Robnik}(1984)}]{Berry_Robnik:1984}%
  \BibitemOpen
  \bibfield  {author} {\bibinfo {author} {\bibfnamefont {M.~V.}\ \bibnamefont
  {Berry}}\ and\ \bibinfo {author} {\bibfnamefont {M.}~\bibnamefont {Robnik}},\
  }\href {\doibase 10.1088/0305-4470/17/12/013} {\bibfield  {journal} {\bibinfo
   {journal} {J. Phys. A: Math. Gen}\ }\textbf {\bibinfo {volume} {17}},\
  \bibinfo {pages} {24413} (\bibinfo {year} {1984})}\BibitemShut {NoStop}%
\bibitem [{\citenamefont {Sakhr}\ and\ \citenamefont
  {Nieminen}(2005)}]{Sakhr:2005}%
  \BibitemOpen
  \bibfield  {author} {\bibinfo {author} {\bibfnamefont {J.}~\bibnamefont
  {Sakhr}}\ and\ \bibinfo {author} {\bibfnamefont {J.~M.}\ \bibnamefont
  {Nieminen}},\ }\href {\doibase 10.1103/PhysRevE.72.045204} {\bibfield
  {journal} {\bibinfo  {journal} {Phys.\ Rev.~E}\ }\textbf {\bibinfo {volume}
  {72}},\ \bibinfo {pages} {045204(R)} (\bibinfo {year} {2005})}\BibitemShut
  {NoStop}%
\bibitem [{\citenamefont {Alhassid}\ \emph {et~al.}(1990)\citenamefont
  {Alhassid}, \citenamefont {Novoselsky},\ and\ \citenamefont
  {Whelan}}]{Alhassid_Whelan:1990}%
  \BibitemOpen
  \bibfield  {author} {\bibinfo {author} {\bibfnamefont {Y.}~\bibnamefont
  {Alhassid}}, \bibinfo {author} {\bibfnamefont {A.}~\bibnamefont
  {Novoselsky}}, \ and\ \bibinfo {author} {\bibfnamefont {N.}~\bibnamefont
  {Whelan}},\ }\href@noop {} {\bibfield  {journal} {\bibinfo  {journal} {Phys.
  Rev. Lett.}\ }\textbf {\bibinfo {volume} {65}},\ \bibinfo {pages} {2971}
  (\bibinfo {year} {1990})}\BibitemShut {NoStop}%
\bibitem [{\citenamefont {Geisel}\ \emph {et~al.}(1991)\citenamefont {Geisel},
  \citenamefont {Ketzmerick},\ and\ \citenamefont
  {Petschel}}]{Geisel_Petschel:1991}%
  \BibitemOpen
  \bibfield  {author} {\bibinfo {author} {\bibfnamefont {T.}~\bibnamefont
  {Geisel}}, \bibinfo {author} {\bibfnamefont {R.}~\bibnamefont {Ketzmerick}},
  \ and\ \bibinfo {author} {\bibfnamefont {G.}~\bibnamefont {Petschel}},\
  }\href@noop {} {\bibfield  {journal} {\bibinfo  {journal} {Phys. Rev. Lett.}\
  }\textbf {\bibinfo {volume} {66}},\ \bibinfo {pages} {1651} (\bibinfo {year}
  {1991})}\BibitemShut {NoStop}%
\bibitem [{\citenamefont {Anderson}(1958)}]{Anderson:1958}%
  \BibitemOpen
  \bibfield  {author} {\bibinfo {author} {\bibfnamefont {P.}~\bibnamefont
  {Anderson}},\ }\href@noop {} {\bibfield  {journal} {\bibinfo  {journal}
  {Phys. Rev.}\ }\textbf {\bibinfo {volume} {109}},\ \bibinfo {pages} {1492}
  (\bibinfo {year} {1958})}\BibitemShut {NoStop}%
\bibitem [{\citenamefont {Zhu}\ \emph {et~al.}(2013)\citenamefont {Zhu},
  \citenamefont {Shao}, \citenamefont {Peng}, \citenamefont {Fan},
  \citenamefont {Huang},\ and\ \citenamefont {Wang}}]{Zhu:2013}%
  \BibitemOpen
  \bibfield  {author} {\bibinfo {author} {\bibfnamefont {L.-H.}\ \bibnamefont
  {Zhu}}, \bibinfo {author} {\bibfnamefont {M.-R.}\ \bibnamefont {Shao}},
  \bibinfo {author} {\bibfnamefont {R.-W.}\ \bibnamefont {Peng}}, \bibinfo
  {author} {\bibfnamefont {R.-H.}\ \bibnamefont {Fan}}, \bibinfo {author}
  {\bibfnamefont {X.-R.}\ \bibnamefont {Huang}}, \ and\ \bibinfo {author}
  {\bibfnamefont {M.}~\bibnamefont {Wang}},\ }\href@noop {} {\bibfield
  {journal} {\bibinfo  {journal} {Opt. Express}\ }\textbf {\bibinfo {volume}
  {21}},\ \bibinfo {pages} {A313} (\bibinfo {year} {2013})}\BibitemShut
  {NoStop}%
\bibitem [{\citenamefont {Abdellatif}\ and\ \citenamefont
  {Kirah}(2013)}]{Abdellatif:2013}%
  \BibitemOpen
  \bibfield  {author} {\bibinfo {author} {\bibfnamefont {S.}~\bibnamefont
  {Abdellatif}}\ and\ \bibinfo {author} {\bibfnamefont {K.}~\bibnamefont
  {Kirah}},\ }\href@noop {} {\bibfield  {journal} {\bibinfo  {journal} {Opt.
  Lett.}\ }\textbf {\bibinfo {volume} {38}},\ \bibinfo {pages} {3680} (\bibinfo
  {year} {2013})}\BibitemShut {NoStop}%
\bibitem [{\citenamefont {Wang}\ \emph {et~al.}(2009)\citenamefont {Wang},
  \citenamefont {Maeda}, \citenamefont {Thomas}, \citenamefont {Takanabe},
  \citenamefont {Xin}, \citenamefont {Carlsson}, \citenamefont {Domen},\ and\
  \citenamefont {Antonietti}}]{Wang:2009}%
  \BibitemOpen
  \bibfield  {author} {\bibinfo {author} {\bibfnamefont {X.}~\bibnamefont
  {Wang}}, \bibinfo {author} {\bibfnamefont {K.}~\bibnamefont {Maeda}},
  \bibinfo {author} {\bibfnamefont {A.}~\bibnamefont {Thomas}}, \bibinfo
  {author} {\bibfnamefont {K.}~\bibnamefont {Takanabe}}, \bibinfo {author}
  {\bibfnamefont {G.}~\bibnamefont {Xin}}, \bibinfo {author} {\bibfnamefont
  {J.~M.}\ \bibnamefont {Carlsson}}, \bibinfo {author} {\bibfnamefont
  {K.}~\bibnamefont {Domen}}, \ and\ \bibinfo {author} {\bibfnamefont
  {M.}~\bibnamefont {Antonietti}},\ }\href@noop {} {\bibfield  {journal}
  {\bibinfo  {journal} {Nature Materials}\ }\textbf {\bibinfo {volume} {8}},\
  \bibinfo {pages} {76} (\bibinfo {year} {2009})}\BibitemShut {NoStop}%
\bibitem [{\citenamefont {Schurig}\ \emph {et~al.}(2006)\citenamefont
  {Schurig}, \citenamefont {Mock}, \citenamefont {Justice}, \citenamefont
  {Cummer}, \citenamefont {Pendry}, \citenamefont {Starr},\ and\ \citenamefont
  {Smith}}]{Schurig:2006}%
  \BibitemOpen
  \bibfield  {author} {\bibinfo {author} {\bibfnamefont {D.}~\bibnamefont
  {Schurig}}, \bibinfo {author} {\bibfnamefont {J.}~\bibnamefont {Mock}},
  \bibinfo {author} {\bibfnamefont {B.}~\bibnamefont {Justice}}, \bibinfo
  {author} {\bibfnamefont {S.}~\bibnamefont {Cummer}}, \bibinfo {author}
  {\bibfnamefont {J.}~\bibnamefont {Pendry}}, \bibinfo {author} {\bibfnamefont
  {A.}~\bibnamefont {Starr}}, \ and\ \bibinfo {author} {\bibfnamefont
  {D.}~\bibnamefont {Smith}},\ }\href@noop {} {\bibfield  {journal} {\bibinfo
  {journal} {Science}\ }\textbf {\bibinfo {volume} {314}},\ \bibinfo {pages}
  {977} (\bibinfo {year} {2006})}\BibitemShut {NoStop}%
\end{thebibliography}
%

\end{document}